\documentclass[10pt,twocolumn,aps,pra]{revtex4}
\pdfoutput=1

\newcommand \be{\begin{equation}}
\newcommand \ba{\begin{eqnarray}}
\newcommand \ea{\end{eqnarray}}
\newcommand \ee{\end{equation}}

\usepackage{float}
\usepackage{epsfig}
\usepackage{epstopdf}
\usepackage{graphicx} 
\usepackage{subfigure}
\usepackage{epsfig,graphics}
\usepackage{epstopdf}
\usepackage{multirow}
\usepackage{url}
\begin{document}
\textheight=10in
\title{
Hydrogen trapping at surface and subsurface vacancies of low-index surfaces of Pd 
}
\author{A. V. Subashiev and  H.  H. Nee}
\affiliation{Target Technology  Company LLC, 564 Wald, Irvine, CA 92618, USA} 

\begin{abstract}
Hydrogen segregation to vacancies in the surface and subsurface layers of (111) and (100) surfaces of Pd is studied in the density functional theory (DFT) approach. Adsorption energies and configurations of various clusters of H atoms at the vacancies are calculated.  The adsorption energy varies  for different sites in the vacancies with the distance to the surface from -0.26 eV (close to that of the bulk clusters) to -0.62 eV. Enhanced binding is found for the sites above the pores produced by vacancies in the subsurface layer. For the (111) surface vacancy the most favorable for segregation are tetrahedral lattice sites, while for (100) octa-sites have higher binding energy. Lattice relaxation effects are  minor for the (111) surface but noticeably enhanced for the (100) surface. Hydrogen segregation to surface layer vacancies is accompanied with minimal charge transfer, which shows itself in cluster configurations.  At high surface coverage the reduction of the cluster formation energy due to the H segregation should result in strongly enlarged concentration of the subsurface vacancy clusters, while the surface remains undamaged due to the prevailing surface adsorption. 
\end{abstract}


\maketitle

\def\dd{\mbox{\small-}}
\section{Introduction}

Interaction of hydrogen with  metal surfaces is of major importance and the topic of active experimental studies for at least two decades. It is due to the major influence of H on the metal mechanical properties (causing embrittlement)  and on chemical properties of the surface and its catalytic activity \cite{Fukai,Gangloff,Greeley}. Besides, it  is of interest for producing materials for H storage and H transparent membranes \cite{Sandrock,Ockwig}.

Adsorption of hydrogen on differently oriented surfaces of the face-centered cubic (fcc) transition metals like Cu, Pd and Ni is well studied both experimentally (using various techniques) and theoretically mainly by DFT calculations. A comprehensive review of the results can be found in \cite{Greeley,Ferrin}. The main result is that the binding energy to all low-index surfaces ranges from 2.3 eV for Au to 3.3 eV for V and is close to 3 eV for Re, Rh, Pd, Ni, and Co.  For evaluation of the equilibrium concentration of H at the surface, one should compare this energy with the binding energy of the hydrogen molecule, which is known to be 4.75 eV, or 2.375 eV per atom. The resulting H adsorption energy for most metals is close to -0.5 eV (here minus reflects binding).  Thus, (i) hydrogen binds to the surface from a molecular phase with dissociation into atomic phase resulting in high coverage of the surface with H atoms at rather low H$_2$ ambient pressure; (2) for moderate values of binding energy it can be desorbed at elevated temperature, which is important for catalytic activity; (3) the adsorbed hydrogen can serve as a source for further penetration into metal to the available octahedral and tetrahedral lattice sites. Penetration of H atoms into bulk interstitial sites generates lattice defects that have rather small formation energy \cite{Nazarov,Lee}  which controls the solubility of H in the metal. The substitution of metal atoms by H requires larger energy and therefore is improbable. Both processes involve H thermoactivation over a barrier (or thermoactivated tunneling through a barrier) of about 0.5 eV or higher. This makes the H loading rate rather low even at high ambient temperature and pressure \cite{Fukai,Morreale}. Electrochemical insertion proved to be more effective, but it creates surface defects that alters the loading mechanism \cite{Benck}.  

Previous DFT calculations \cite{Nazarov,Lee,Vekilova,Lischka,Lober} showed that the formation energy for the H atom in a Pd interstitial octahedral site is around -0.11 eV, so that penetration of H to these sites is a favorable process. This unique property of Pd (negative solution energy for H atoms) result in very high H solubility in a metallic phase.  The activation energy for H to enter into the subsurface layer is 0.4 eV,  and 0.3 eV for entering the second subsurface layer, while the activation energy for the bulk diffusion is only 0.23 eV \cite{Ferrin}. Therefore the kinetics of Pd charging with H is controlled by the escape rate from the surface into the bulk.  At high loading there are competing processes: (i) formation of the PdH$_x$ hydride phase, and (ii) generation of extra vacancies with fast hydrogen segregation at them and further ordering in approaching the equilibrium state. Energy gain in the binding  of several H atoms compensates the system energy loss in the vacancy generation. As a result  superabundant vacancy (SAV) phases, (e.g. Pd$_3$VacH$_4$) can emerge \cite{Fukai_Oku,Fukai_SAV,Zhang,Buckonte}.    

In addition to the numerous experimental studies and DFT modeling of energetics of H interaction with low index surfaces of Pd (see the reviews  \cite{Greeley,Ferrin} for the references), and equally comprehensive studies of H interaction with vacancies in the metal bulk \cite{Nazarov,Lee,Vekilova,Lischka} and thermodynamics of the SAV state \cite{Fukai,Zhang,Kirchheim}, the advances in {\em ab-initio} molecular dynamics (AIMD) made possible investigation of kinetics of H$_2$ adsorption on clean and pre-covered Pd low-index surfaces \cite{Gross10}.

These studies showed that the H$_2$ dissociative adsorption processes is sensitive to the Pd surface "poisoning", including the surface coverage by the adsorbed H atoms: in the so-called "site blocking picture" for the H$_2$ dissociative adsorption at least the two adjacent empty sites on the H populated surface should be available. However, the experimental studies of the Pd (111) surface by the scanning tunneling microscope \cite{Mitsui1} demonstrated that the divacancy type of pairs at the H covered surface remain inactive and more complex structures that include at least 3 surface vacancies are most active in adsorption-desorption processes. Such structures are easily formed due to a rapid H vacancy diffusion at the surface.  

Further AIMD modeling \cite{Gross11} showed a more complex picture that suggests the competition of several mechanisms, including the H$_2$ dissociative adsorption at the single vacancy in the H overlayer with filling the bridge site on the Pd(100) surface.

Note that the native defects on the nominally clean surface such as steps, adatoms, and vacancies can modify both the kinetics and the thermodynamics of Pd hydrogen absorption acting like active sites or the H traps. The detailed information about these processes is important for interpretation of the properties of attainable surfaces, experimental studies of spectra of temperature programmed desorption, for engineering catalytic properties of Pd and so on. 

The formation energy of a single vacancy on the metal surface is considerably smaller than for the bulk \cite{Kauffman}, which should facilitate formation of clusters of hydrogen segregated at these vacancies. The H insertion energy into the subsurface layers also depends on the distance from the surface, so the H binding to a vacancy becomes dependent on the H position relative to the surface. These factors show themselves in the mechanisms of H penetration from the surface into bulk. In addition, the lattice relaxation effects at the surface are much more pronounced and influence both the vacancy formation, H interstitial state and H segregation to the vacancies.  

Here we report the studies of the H interaction with vacancies in the surface and subsurface layers in Pd using density functional theory (DFT) modeling. We show that the subsurface vacancies should be important for H charging kinetics and SAV state formation.

\section{Methods}
\subsection{Computational Details}
Calculations were performed using Quantum Espresso  (QE) \cite{QE} {\em ab-initio} simulation package with the generalized gradient approximation (PBE version \cite{PBE}) for exchange-correlation functional. We used ultrasoft pseudopotentials \cite{UPF} for both Pd and H atoms. For these calculations one has to start with the choice of suitable supercells for (111) and (100) surfaces, that enables the studies of these effects with reliable accuracy. 
 
The effects at the surface were modeled using a periodic structure with a supercell consisting of a metallic slab separated from its periodic images by a vacuum layer of a thickness of $\approx 15 $ \AA. For calculations of the surface and subsurface vacancies and also $n$H-V clusters (with $n$=1, 2,...,$n_{max}$),  4-layer structures with a 36-atom supercell for  (111) surface (see Fig. \ref{SurVac111-3h3} and a note \cite{PP}) and a supercell with 32 atoms in a cell for (100) surface (shown in Fig. \ref{5Hat100surv}) were used.  One missing atom in the surface layer of the 8-atom cell for the (100) and of the 9-atom cell for the (111) surface  resulted in 1/8 (or 1/9) surface coverage by the defects. It is sufficient to eliminate the effects of interaction between the defects and is smaller than the coverage used in the previous studies \cite{Lischka,Lober,Loovik}. A k-point Monkhorst-Pack grid 8x8x1 is known to be adequate for the  selected supercells \cite{Singh-miller,Kauffman} which was checked in the surface energy calculations.  To estimate convergence of the surface energy we used supercells with 4 atoms at the surface and an increasing number of layers in the slab from 4 to 10 and then compared the results with that of the 4-layered 32-at and 36-at supercells. The variations within 10 meV were considered to be satisfactory. 
 
In calculations of bulk parameters we used the 27-at, 32-at  and 108-at supercells without the vacuum layer and a  16384 k point $\times$ atoms mesh.  The cutoff energies were chosen to obtain the accuracy of energy scaling  to be within 2 meV/atom when going from 4-at to a 27-at, to a 32-at, and further to a 108-at supercell. Plane-wave cutoff energy was set at  30 Ry  and the energy cutoff for the electron density was maintained at 360 Ry, which provided convergence of the results for the ultrasoft pseudopotentials. The  Marzari-Vanderbilt \cite{M-V} smearing was used with the smearing energy set at 0.l eV. 

The most critical issue in comparing various configurations of segregated H atoms at the vacancies is achieving close to total lattice relaxation in the final state. Therefore lattice relaxation to equilibrium was reached when the forces on each atom were below 5 meV/\AA. The estimated accuracy of the results ($\approx$ 0.1 eV ) is close to that of the Refs. \cite{Ferrin,Greeley,Nazarov}, see \cite{SM}.
 
To ensure reliability of the results, we have calculated the fcc lattice parameter ($a=3.93$ \AA), bulk elastic modulus, the cohesion energy and vacancy formation energy of Pd \cite{SM}.  Good agreement with previously published results indicates the relevance of the DFT calculation parameters. 
 
Obtained values for the vacancy formation energy $E^f_{PBE}(V)$= 1.2 eV (PBE) and $E^f_{PZ}(V)$=1.44 eV (LDA)  are close to the calculated GGA and LDA values from Refs. \cite{Nazarov,Vekilova} (see \cite{SM}).  
It is known that for most metals GGA approximation gives the vacancy energy values smaller than experimentally observed ($\approx$ 1.5-1.7 eV) and smaller than that obtained using LDA approximation. The discrepancy is due to the shortcomings of both GGA and LDA in accounting for the large variation of the electron densities inside vacancies.  They can be minimized using empirical correction for the internal surface energy of the vacancies \cite{Carling,Nazarov2}. 

The other approach would be to use more sophisticated modern exchange-correlation functionals designed for more accurate evaluation of surface effects, such as PBEsol with the first-principles gradient expansion for exchange energy \cite{PBEsol} or HSEsol with further modified description of the short-range exchange energy \cite{HSEsol}. Even more promising is the use of meta-GGA  functionals with the improved account for the density of electron orbital kinetic energy \cite{Sun,Pedrew_W}. Comparison of calculations of basic Pd parameters for PZ,PBE, and PBE-sol functionals and corresponding US pseudopotentials as well as the results for the TPSS meta-GGA functional from Refs. \cite{Sun,Medasa} is given in \cite{SM}.  

We have found that PBE approximation gives satisfactory results for the surface energy and H binding energy to the surface, in line with the former results, see below. LDA approximation strongly overestimates hydrogen binding energy to the vacancies (in agreement with former observations for the  surface binding \cite{Greeley,Loovik} and binding to the bulk vacancies \cite{Nazarov2}). Therefore here we present the results obtained with PBE functional. The main results were further checked using PBE-sol functional and US pseudopotential, they are marked as R$^{(s)}$ or noted separately.

As we are primarily interested in variation of H binding to the vacancies with the vacancy layer depth and the H ambient concentration,  we  have not implemented posterior corrections to the vacancy formation energies.  
  
Hydrogen adsorption energy  $E_{abs}$  from a gaseous phase is calculated through a difference between the energy of a slab supercell with an adsorbed H atom $E_{slab}(N_{sc},{\rm H})$ and the energy of the slab without the adsorbent  $E_{slab}$ as \cite{Loovik}
\be
E_{abs}=E_{slab}(N_{sc},{\rm H})-E_{slab}(N_{sc}) - 1/2 E(H_2) 
\label{adso}
\ee
where $E(H_2)$ is the energy of a free H$_2$  molecule. This energy is calculated in QE as an energy per cell in a crystal of H$_2$ molecules with a very large lattice constant (10 -15 \AA).  Similar equations are applied for evaluation of the solution energy to octahedral and tetrahedral sites in the bulk, and for calculation of the H adsorption to vacancies. 

Also of interest is the hydrogen binding energy $E^b({\rm V}; n {\rm H})$  in a cluster with $n$ H-atoms segregated to a vacancy V. It can be evaluated through the total energy of four different supercells with $N$-sites: (i) with no defects, $E_t(N)$, (ii) with interstitial H, $E_t(N,{\rm H})$, (iii) with a vacancy $E_t(N-1)$ and (iv) with both vacancy and $n$ segregated H atoms, $E_t(N,n {\rm H})$,  using the following formula \cite{Wolv} :  
\ba \label{TrapTang}
E^b({\rm V}; n {\rm H})  =  E_t(N-1,n {\rm H})- E_t(N-1) \\ \nonumber -n[E_t(N,{\rm H})- E_t(N)]
\ea
According to  Eq. (\ref{TrapTang}) the binding energy can be viewed as a difference  between the energies of absorption of hydrogen atoms to the positions in a vacancy cluster and absorption to an interstitial site. It can also be considered as an infinite separation energy (ISE), i.e., a difference between the  energy of the decorated vacancy cluster and $n$ bulk Pd cells and the energy of  the "infinite separation" state of $n$ cells with an interstitial H atom and a cell with a vacancy defect  \cite{Wolv}.  One can apply the ISE approach, i.e. Eq. (\ref{TrapTang}), for a cluster at the surface or subsurface layer,  inserting  calculated energies for a slab of finite thickness. Note that Eq. (\ref{adso}) is also based on the ISE calculation, and that listed in \cite{Ferrin}  the H binding energies to the surface for various metals are  the ISE's relative to the energy of atomic hydrogen.

For an interstitial H atom starting from the third layer the insertion energy (in the square bracket of Eq. (\ref{TrapTang})) is very close to that of the bulk, and we will not consider the difference between the H binding to a surface vacancy from the bulk and binding from the layers close to the surface. 

Calculation of insertion energy to the Pd bulk gives $E_{i,b}=-0.11$ eV for octa- sites and -0.09 eV for the tetra-sites including zero-point energy corrections \cite{SM}. Therefore, in our calculations the  binding (or segregation) energy for a $n$H-V$_i$, $i =s, ss, b$ cluster is smaller than adsorption (or absorption) energy by -0.11 eV. Below we present the results for absorption energy, which also reflects H segregation from the bulk Pd.  

At low external pressure and temperature $T$=0 the  DFT-calculated adsorption and absorption energies using Eq. (\ref{adso}) coincide with the defect formation energies.  The equilibrium concentration of the defects (including concentration of various $n$H-V$_s$ and  $n$H-V$_{ss}$ clusters with $n$= 0, 1, 2, .. $n_{\rm max}$) at a given temperature and pressure depends on the enthalpy of the defect formation $H_f^i(P,T)$, which includes the terms with the equilibrium chemical potentials \cite{Nazarov,Kirchheim}. It can be written as
\be \label{entha}
H_f^i = E_f^i-\Delta n_{Pd}^i \mu_{Pd} - \Delta n_{H}^i \mu_{H}(P,T) 
\ee
Here $E_f^i$ is the defect formation energy, index $i$ labels the sort of the defect, $\Delta n^i$ is the change in the number of atoms in the defect,  including H atoms segregated at the surface, for which $\Delta n_{Pd}^i = 0 $, and $n_{H}^i$=1. For a vacancy $\Delta n_{Pd}^i = -1 $, and  $\mu_{\rm H}(P,T)$ and $\mu_{\rm Pd}(P,T)$ are the H and Pd chemical potentials; for $n$H-V$_i$ clusters $\Delta n_{H}^i=n$.
\begin{table*}[t]
\begin{caption}{\label{surf} Surface energies $\sigma$, and vacancy formation energies for surface, $E_{v,s}$, subsurface, $E_{v,ss}$ vacancies (in eV) at clean low-index Pd surfaces, and $E_{v,b}$ for the bulk Pd, calculated using GGA (PBE, PBE-sol) and LDA (PZ) functionals and ultrasoft pseudo-potentials; also shown are results of experiments  \cite{Tyson}$^a$, \cite{Kraftmak}$^b$, \cite{Voter}$^c$ and previous calculations\cite{Singh-miller}$^d$, \cite{Kauffman}$^e$, \cite{erem}$^f$,\cite{Nazarov}$^g$,\cite{Vekilova}$^h$.}   
\end{caption}
\begin{centering}
\begin{tabular}{llllllllllll}
\hline
 -  &          & (111)  &    &    &  &           &         & (100) & \\ 
\hline
  -        & PBE   & PBE-sol & PZ  &  exper    & others            & & PBE  & PBE-sol  & PZ  & exper & others  \\
$\sigma$   & 0.58  & 0.67   & 0.72 &  0.82$^a$ & 0.56$^d$,0.68$^e$ & & 0.73 & 0.9  & 0.97  & 0.82$^a$ & 0.74$^d$ \\
$E_{v,s}$  & 0.93  & 1.17   &1.06 &  - & 1.03$^e$, 0.78 $^f$       & & 0.61  & 0.67  &  0.72 & \ \ - & 0.57$^f$  \\
$E_{v,ss}$ & 1.22  & 1.27 & 1.36  &   -  &  1.24$^e$, 1.44$^f$ & & 1.2 & 1.25 & 1.32 &  \ \ -  & 1.44$^f$  \\
$E_{v,b}$  & 1.23  & 1.34 & 1.51  & 1.51$^b$,1.54$^c$  &  1.2$^g$, 1.37$^f$, 1.44$^h$ & & &   &   &   \\ 
\hline
\end{tabular}
\end{centering}
\end{table*}

For the surface layer in equilibrium to the crystal bulk the  variation of $\mu_{\rm Pd}(P,T)$ is negligible and $\mu_{\rm Pd}(P,T) \approx E_{pa}$ where $E_{pa}$ is the total (negative) energy of Pd crystal (i.e. supercell) per Pd atom. For the hydrogen $\mu_{\rm H}(P,T)$=1/2$\mu_{\rm H_2}(P,T)$. In H$_2$ gaseous atmosphere $\mu_{\rm H_2}(P,T)$ can vary in a range of about 1 eV, from negative values to positive; tables for $\mu_{\rm H_2}(P,T)$ can be found in \cite{Fukai}. For gaseous phase $\mu_{\rm H_2}(P,T)$ can be  written as
\be \label{mu}
\mu_{\rm H_2}(P,T) =\mu_{\rm H_2}(P_0,T)+ k_B T \ln \frac{P}{P_0} 
\ee
Here $\mu_{\rm H_2}(P_0,T)$ is the chemical potential of H$_2$  at a pressure $P_0$. It includes the contribution of ideal-gas entropy, and contributions from rotations and vibrations of the H$_2$ molecule. At $T=T_0$=300 K and $P_0$=1 bar $\mu_{\rm H_2}(P_0,T_0)$=-0.31 eV. The  last term in Eq. (\ref{mu}) shows the variation of $\mu_{\rm H_2}(P,T)$ with pressure at a temperature $T$.

Concentration of  different defects can be found from the minimum of the Gibbs energy of the system with respect to the number of particles in various positions \cite{Monast,NazarovFe,Ji}. The resulting distributions  depend on the ratios $H_f^i /(k_B T)$ and resemble Fermi-Dirac statistics: the occupation of the available lattice (or surface) sites is exponentially small at $H_f^i >> k_B T$ and approaches unity when the formation enthalpy becomes negative (e.g. due to H segregation). Correct occupation numbers in this limit should take into account interaction between the defects which will modify the equation for the defect formation enthalpy.

\subsection{Properties of  (111) and (100) surfaces of Pd}
Results of the calculations of the Pd surface properties are summarized in the Table \ref{surf}, together with experimental data and the results of previous calculations. As one can observe, (i) calculated surface and vacancy energies are underestimated in GGA and overestimated in LDA (PZ) approximation, (ii) obtained results are in close agreement with former DFT calculation results but differ from the embedded atom theory (EAT) results of \cite{erem}. The most notable difference is the EAT prediction of enlarged formation energy for subsurface vacancies, while DFT calculations give a smooth increase of the vacancy formation  energy  with the distance from the surface. We are unaware of experimental studies of this effect.

One can observe that: (i) the formation energies obtained using LDA approximation are  regularly somewhat higher than that obtained with PBE functional;  (ii) the surface formation energy is lower for the (111) surface of Pd while the surface vacancy formation energy is lower for the (100) surface. The origin of these difference can be interpreted in terms of the broken bonds between Pd atoms in the creation of the defect.  Qualitatively these data should be compared to  cohesive energy for the Pd lattice which is 3.9 eV per atom, see \cite{SM}. Every Pd atom has 12 nearest neighbors, which make  the largest contribution to the cohesive energy. In a (100) surface layer a surface atom has 4 nearest neighbors missing (compared to the bulk), in a (111) surface a surface atom loses 3 nearest neighbors.  Therefore the surface energy should be larger for the (100) surface, which correlates well with the results presented in Table \ref{surf}. Using the same arguments one can interpret the difference in creating bulk and surface vacancies through the ratio of broken bonds (the surface vacancy formation energy is expected to be higher for a (111) surface).  Consequently, one could argue that the difference between the vacancy creation energy in the subsurface layer and the bulk should be less pronounced and should be attributed to elastic energy relaxation effects. Apparently, the lattice relaxation effects are more important for the (100) surface (see below). 
\subsection{Adsorption of H  on Pd surfaces}

\begin{figure}[b]%
 		\centering
		\subfigure[]{{\includegraphics[width=6.6cm]{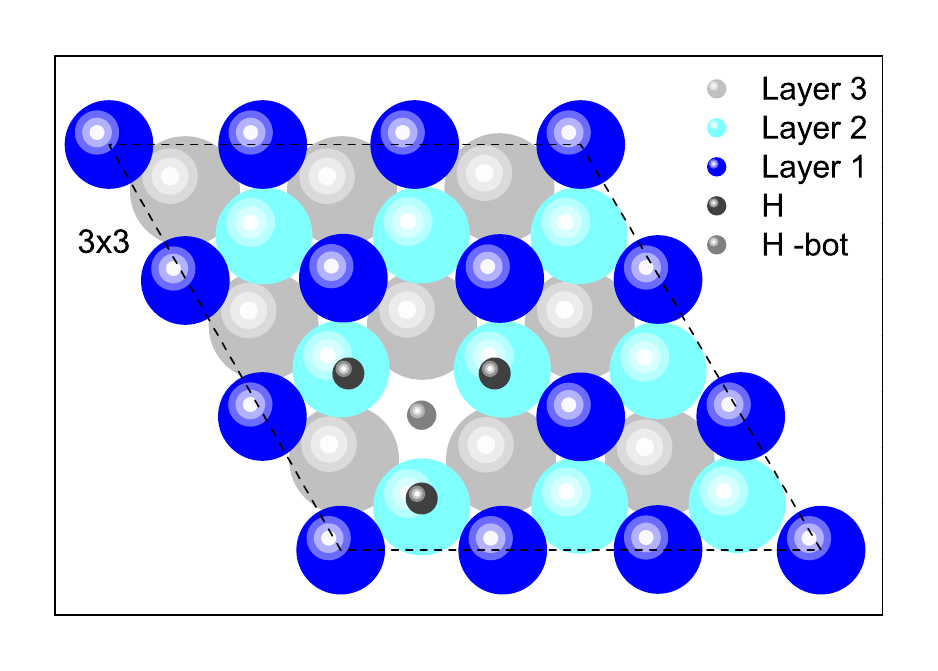} }}%
		\qquad
    \subfigure[]{{\includegraphics[width=6.6cm]{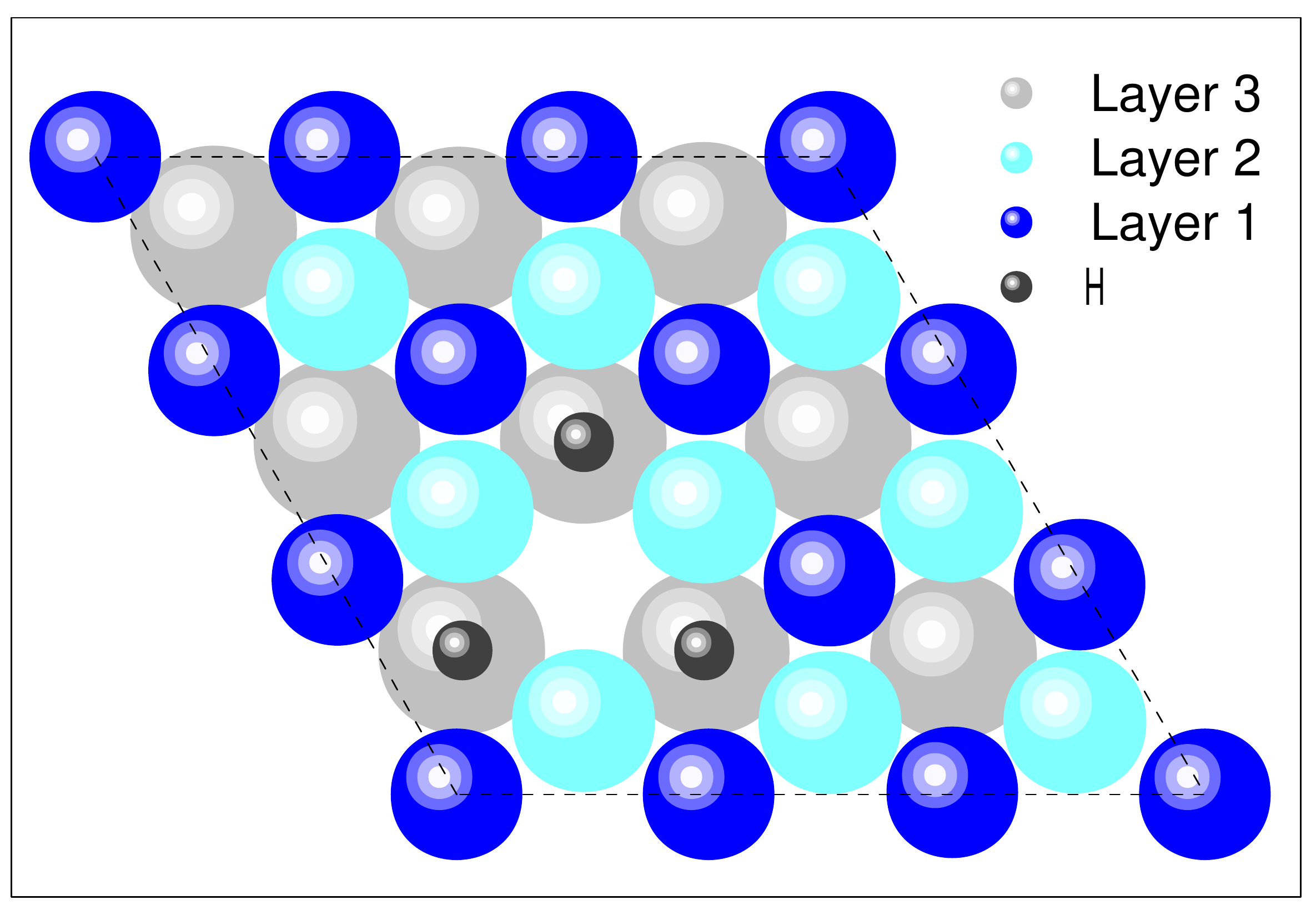} }}%
    \caption{\label{SurVac111-3h3} Segregation of hydrogen  atoms (small dark balls) to the surface vacancy in Pd (111)  surface.  Surface layer is shown by blue balls, subsurface atoms are light-blue; the next layer atoms are painted gray. The surface unit cell is shown by the dashed line. H atoms above the second layer (panel a) are at tetrahedral sites of Pd lattice (below "hollow" hcp sites of the surface); the central H atom is at the bottom tetrahedral site, atoms above the third layer (panel b) are at octahedral sites, below fcc sites at the surface.}
	\end{figure}

There are several high-symmetry sites on a clean surface that are preferable for adsorption: at the top above a Pd atom, at the "bridge" between two nearest Pd atoms, or "hollow" sites above the center of a triangle of Pd atoms at the (111) surface or above the center of a square of Pd atoms on the (100) surface.  It is known \cite{Ferrin} that  the H$_2$ molecule is dissociatively adsorbed  at the Pd surface at the "hollow" sites from a low to high surface coverage.  In all cases the most important parameters are the adsorption energy and the distance $\delta d$ \ between the adsorbed atom and the Pd surface plane.   
The calculated hydrogen adsorption energies on the (111) surface are slightly different for two hollow adsorption sites. For fcc sites (with Pd atom below H atom), E$_{abs, fcc}$=-0.57 eV; for the hexagonal close packed lattice (hcp) sites, above the tetrahedral pore between Pd atoms (see also Fig. \ref{SurVac111-3h3}), E$_{abs, fcc}$=-0.55 eV. These values are very close to the values (E$_{abs, fcc}$=-0.6 eV, and E$_{abs, fcc}$= -0.56 eV)  in Ref \cite{Greeley} (without corrections on H ZPE energy at the surface, that lowers these values). The distances $\delta d_{fcc} = 0.8$  \AA \ for fcc site and $\delta d_{hcp} = 0.68$  \AA, are close to former findings \cite{Loovik,Ferrin,Greeley}. The experimental data for adsorption  energy (from desorption experiments) give a smaller value, close to -0.45 eV \cite{Felter}  and a very close distance $\delta d=0.8 \pm 0.1$ \AA; the deviation may be a result of the difference in coverage and the lattice relaxation state. 
		
For the (100) surface the calculated hydrogen adsorption energy is E$_{abs, (100)}$=-0.48 eV, close to the  experimental data E$_{abs, (100)}$=-0.47 eV  \cite{Lischka}, while the estimated distance to the surface is  $\delta d=0.4$ \AA. 

Thus, the reliability of our calculations  is confirmed by comparison with available data of former publications \cite{Singh-miller,Kauffman,erem} both for the bulk and surface properties as well as the data for H insertion and binding to vacancies in the bulk \cite{Nazarov,Lee,Morreale}. 

\section{Hydrogen segregation to surface and subsurface vacancies}
\subsection{H adsorption on surface and subsurface vacancies at clean Pd surfaces}

Now we discuss hydrogen segregation to various $n$H-V$_s$ and $n$H-V$_{ss}$ clusters with $n$=1, 2,..$n_{max}$ calculated within GGA approximation. There are numerous combinations of sites available for segregation into the clusters. In subsequent segregation the binding energy for a chosen $n$ depends on the site population. Similarly to H segregation to vacancies in the bulk Pd \cite{Vekilova,Nazarov} there is a repulsion between segregated H atoms, that reduces the segregation energy of a pair of H atoms to the nearest sites, so  population of the opposing sites is preferable and the subsequent segregation energy decreases with the population of available sites. This repulsion is rather small for the sites of one type (e.g. octa- or tetra-sites) but eliminates configurations in which both types of sites are occupied. The other factor is the distance from the site to the surface layer.  Here we present the results for the symmetrical clusters with maximal population of equivalent sites, which allows us to estimate the segregation energy per H atom for the site of a given type.   

We start with the surface and subsurface vacancies at the Pd (111) surface. Total adsorption energies $E_{tot}$  and the average adsorption energies $E_{av}$ per H atom for exemplary $n$H-V$_s$ clusters are presented in Table \ref{s111}.
Adsorption energy to the bottom tetra-site of the surface vacancy [see Fig. \ref{SurVac111-3h3} (a)] is E$_{s, bot}$= -0.34 eV, which is close to  E$_{abs, tet.bulk}$, consistent with the similar geometry of surrounding Pd atoms. The distance to the surface $\delta d=0.65~l$ is somewhat smaller than the distance $0.75~l$ from a tetrahedral site to a vacancy center (here $l$ is the distance between the atomic layers), reflecting lattice relaxation effects. 

Segregation of 3 H atoms to 3 equivalent subsurface tetrahedral sites gives the adsorption energy $E_{s, tetra}$= -0.46 eV per atom, which is smaller than for hcp sites at the surface, but higher than that for the bottom state. The distance from the surface layer to H atoms $d=0.17 l$  is also smaller than the distance $0.25 l$ to the tetrahedral sites. Population of 3 octahedral sites (see (b) panel of Fig. \ref{SurVac111-3h3}) gives smaller adsorption energy per atom  $E_{s, octa}$= -0.36 eV, with the distance to  the surface $\delta d=0.33 ~l$, while initial distance is $d=0.5~ l$. 
\begin{table}[tb]
\begin{caption}{\label{s111} Adsorption energies (in eV)  and distances from the surface atomic layer in units of inter-layer distance $l=a/\sqrt{3}$  for H to surface vacancy clusters $n$H-V$_s$  at Pd (111) surface and 1/9  coverage.}
\end{caption}
\begin{tabular}{llllll}
\hline
 n & positions & $E_{tot}$ & $E_{av}$  & $\delta d_{top}$  & $\delta d_{bot}$ \\ \hline
 1 & bot   & -0.34 & -0.34   & - &  0.65  \\
 3 & tetra  & -1.37 &  -0.46 &  0.17  & - \\
 3 & octa   & -1.07  & -0.36  &  0.33  & -   \\
 4 & tetra bot & -1.44 & -0.36 & 3 x 0.15 & 0.77  \\
 \hline
\end{tabular}
\end{table}

Finally, segregation of 4 H atoms to all available tetrahedral sites gives the average adsorption energy  $E_{s, tetra}$= -0.361 eV, smaller than the sum of $E_{tot}$ for 1H-V$s$ and 3H-V$s$ clusters. Besides the position of the H atom at the bottom is shifted below the tetrahedral site, and the other 3 H atoms are shifted up, closer to the surface.     

As seen from Table \ref{s111} the total adsorption energy for 3H-V$s$ and 4H-V$s$ clusters exceeds the surface vacancy formation energy, so that these clusters should be stable and could initiate the formation of excessive $n$H-V$s$ clusters in the thermal equilibrium state. However, the occupation of tetra-sites by H atoms eliminates the possibility of H adsorption to the surface sites above them. Therefore the total negative formation energy will be obtained only in the case of the clean surface.
\begin{table*}[tb]
\begin{caption}{\label{subs111} Adsorption energies (in eV)  for H to vacancy clusters $n$H-V$_{ss}$  in the subsurface layer at Pd (111) surface and 1/9  coverage, the distances to H sites from the surface layer $\delta d_{top}$ and  $\delta d_{bot}$ are in the units of interlayer separation $l$.}
\end{caption}
\begin{tabular}{llllll}
\hline
 n & positions & $E_{tot}$ & $E_{av}$  & $\delta d_{top}$  & $\delta d_{bot}$ \\ \hline
1 & pore    &   -0.52 (-0.47$^{(s)}$) & -0.52 (-0.47$^{(s)}$)  & -0.33 &  -   \\ 
1 & tetra bot   & -0.34 & -0.34   &  - &  1.65    \\
 2 & tetra up-bot  & -0.75 & -0.37 &  0.33 & 1.66 \\
 4 & tetra  & -0.98 & -0.24 & 3x 1.25  & 1.8 \\
 6 & octa-tetra  & -1.70  & -0.28  & 3x 0.63 & 3x 1.29  \\
 8 & tetra   & -2.04  & -0.25  &  0.07, 3x0.63 & 3x1.28, 1x1.8   \\
  \hline
\end{tabular}
\end{table*}

We also considered hydrogen segregation into $n$H-V$_{ss}$ at the vacancy in the Pd (111) subsurface  layer. The results are presented in Table \ref{subs111}. First, we note that in addition to filling available tetrahedral and octahedral sites, there is an adsorption site above the subsurface vacancy site and above the surface, that we denote as a "pore" state in the Table \ref{subs111}.  This state has the adsorption energy E$_{ss, pore}$= -0.52 eV, 20 meV less than for the initial hcp "hollow" state (see Table \ref{surf}), and is shifted by the same distance $d=0.8$ \AA \ above the surface.  Absorption energy calculated using PBE-sol functional (shown in  parentheses) give a somewhat smaller energy, but again close to that for the initial hcp "hollow" sites. In addition, there are the sites available for segregation as in the bulk, which are modified by the presence of the surface. Among them though less modified is segregation of H to a tetrahedral site at the bottom of the cluster. The adsorption energy  E$_{ss, bot}$= -0.34 eV, is the same as in 1H-V$_s$ cluster as well as the distance to the vacancy atomic layer.  
\begin{figure}[b]
\includegraphics[width=6.8cm]{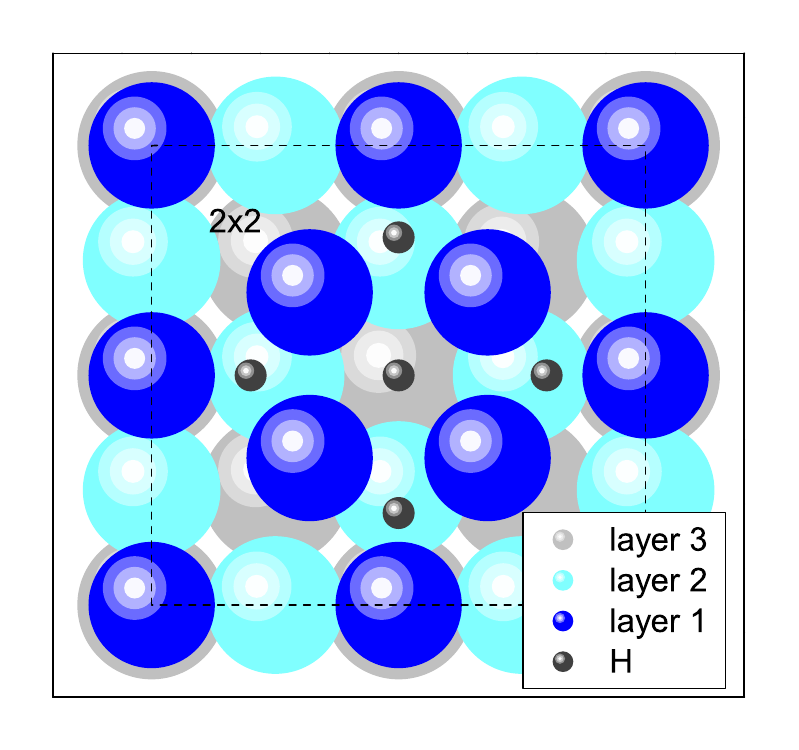} 
\caption[]{\label{5Hat100surv}(color online) Segregation of 5 hydrogen  atoms (small dark balls) to the surface vacancy in Pd (100)  surface; surface layer is shown by blue balls, subsurface atoms are light-blue; atoms of the next layer are painted gray, dashed line shows the surface cell. All H atoms are shifted from octahedral sites of Pd lattice  (4 "hollow" sites of the surface and a bottom site in the subsurface layer, see Table \ref{s100}).}
\end{figure}

Surprisingly, 4 H atoms segregated to tetrahedral sites below the vacancy layer have unshifted distances to the surface but lowered average adsorption energy. It is increased in a 6H-V$_{ss}$ cluster  with 3 atoms in tetra-sites below the vacancy plane and 3 above this plane, see Table \ref{subs111}. For a cluster 8H-V$_{ss}$ with all occupied tetrahedral sites the average adsorption energy is E$_{ss, avr}$= -0.25 eV.  A similar decrease of the average segregation energy with the increase of the number of segregated H atoms was found for the bulk vacancies in Pd \cite{Nazarov}.  Again, importantly, the subsurface vacancy creation energy is lower than for the bulk, and the total segregation energy starts to exceed it for $n \ge 5$; so that clusters with a high number of H atoms can be easily accumulated in equilibrium conditions. 

Atomic structure of the surface and subsurface layers near a  vacancy at the Pd (100) surface is shown (for the case of the 5H-V$s$ cluster) in Fig. \ref{5Hat100surv}. Results of calculations for $n$H-V$s$ clusters at  the (100) surface are presented in Table \ref{s100}.
\begin{table}[t]
\begin{caption}{\label{s100} Adsorption  energies  (in eV)  for H  to surface vacancy clusters $n$H-V$_s$ for a vacancy at  Pd (100) surface at 1/8  coverage, and distances from H atoms to the surface layer plane (in units of interlayer distance $l=a/2$). }
\end{caption}
\begin{tabular}{llllll}
\hline
 n & positions & $E_{tot}$  & $E_{av}$ &  $\delta d_{upp}$ & $\delta d_{bot}$  \\ \hline
 1 & octa-bot   & -0.33  &  -0.33  & - &  0.45  \\
 1 & octa-side  & -0.47  & -0.47  & -0.02 &    \\
  2 & octa   &  -0.96   & -0.48  & 0.0   & - \\
 3 & 2 octa, 1 bot   &  -1.25   & -0.42  & -0.02   & 0.5 \\
  4 & 4 octa    &  -1.98  & -0.5  & -0.03   & -   \\
 4 & 4 tetra  &  -0.99  & -0.25  & -   & 0.23  \\  
 5 & 5 octa    &  -2.22  & -0.44 & -0.02  & 0.54   \\
\hline
\end{tabular}
\end{table}

Adsorption energy to the bottom octahedral site at the subsurface layer of the surface vacancy is E$_{s, bot}$= -0.33 eV, higher than for a vacancy in the bulk and close to  E$_{abs, tetra}$ for the bulk vacancy. The position of the H atom is shifted up to the surface by 0.05$~l$. Adsorption energy  of 1 H atom to the octahedral site close to the surface layer E$_{s, octa}$=-0.47 eV, are smaller than for  "hollow" sites at the (100) surface, but still enlarged  if compared to subsurface binding or compared to that in the bulk. For 2 H atoms at the surface sites close to the vacancy the adsorption energy is further enlarged. In the case of 2 H atoms at the surface sites and one at the vacancy bottom the total adsorption energy is only 43 meV smaller than the sum of the energies for 1 bottom H and a pair at the surface, so the  repulsion between segregated H atoms at the surface vacancy is quite weak. The same conclusion can be drawn comparing the binding energies for 5 segregated H atoms and 4 H atoms (the difference is slightly smaller than the adsorption energy for the bottom site). The adsorption energies of 4 H atoms to  tetrahedral sites below the surface layer appeared to be smaller than for the bottom octahedral site and for the bulk, apparently due to repulsion between the H atoms. 
\begin{table*}[tb]
\begin{caption}{\label{ss100} Adsorption energies (in eV)  for H  to  vacancy clusters $n$H-V$_{ss}$  for vacancy in  the second layer at Pd (100) surface and 1/8 coverage. Presented are total energy, average energy per atom and distances from H atoms to the surface layer plane (in units of interlayer distance, $a/2$). }
\end{caption}
\begin{tabular}{llllll}
\hline
 n & positions & $E_{tot}$  & $E_{av}$ &  $\delta d_{upp}$ & $\delta d_{bot}$  \\ \hline
1 & pore   & -0.72 (-0.62$^{(s)}$)   &  -0.72 (-0.62$^{(s)}$)   & -0.11 &  -   \\
1 & bot   & -0.33  &  -0.33      &  & 2.12   \\ 
2 & pore-bot   & -1.00 (-0.86$^{(s)}$) &  -0.50 (-0.43$^{(s)}$)  &  -0.12 & 2.12   \\ 
2 & octa   &  -0.6   & -0.3  & 0.88  & - \\
3 & 2 octa, 1 bot    &  -0.89   & -0.29  & 2x 0.89 &  2.16 \\
4 & 1 pore 2 octa, 1 bot    &  -1.51   & -0.38  & -0.085, 0.45&  2.16 \\
5 & 4 octa, 1 bot  & -1.47 & -0.29   & 0.44  & 1.05  \\
6 & 1 pore 4 octa 1 bot &  -2.05 (-1.79$^{(s)}$)   & 0.34 (-0.3$^{(s)}$) & -0.114, 0.88  &  2.16 \\
8 & 8 tetra  &  -2.16 & -0.27  & 0.29   & 1.48 \\  
\hline
\end{tabular}
\end{table*}

\begin{figure}[b]
\includegraphics[width=6.8cm]{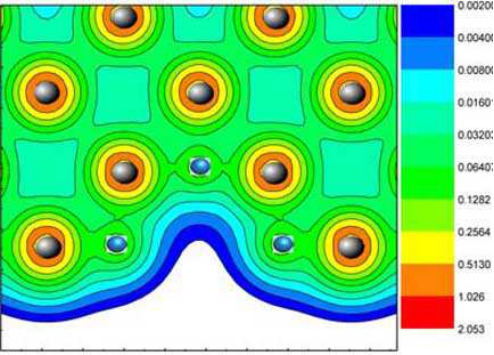} 
\caption{\label{3H-Rhoat100surv}Contour plot of the charge distribution in the plane of segregated 3 H atoms for 3H-V$_s$ cluster at Pd (100) surface; one H atom is at a vacancy bottom octahedral site, 2 H are at the surface "hollow" sites nearest to the vacancy, as in Fig \ref{5Hat100surv}.  Color panel displays density values in atomic units (1$/$au$^3$=6.748 e$/$\AA$^3$);  contour line spacing is in log scale with an increase by a factor of 2 to the next contour line. Gray balls show Pd core regions, small blue balls correspond to H atoms.}
\end{figure}

The electronic density distribution for 3H-V$_s$ cluster with 1 H at the bottom and 2 H atoms in the opposing octa sites at the surface in a plane of segregated H atoms is shown in Fig. \ref{3H-Rhoat100surv}. While the H state at the vacancy bottom is similar to that in the vacancy in the bulk, the surface layer H atoms are placed in the region of small electronic density and move closer to the Pd atoms which results in higher binding energy. 

The lattice relaxation effects manifest themselves in the positions of the segregated H atoms in the surface layer, see Figs. \ref{5Hat100surv}, \ref{3H-Rhoat100surv} for the  5H-V$_s$ and 3H-V$_s$ clusters. For both clusters one can observe the shift of H surface atoms out from the vacancy and the shift of surrounding Pd atoms to the vacancy, while the bottom H atom is slightly shifted below the subsurface layer. The shift of Pd atoms to the surface vacancy in the absence of the segregated H atoms is 2 times smaller.

The results for the vacancy in the subsurface plane of the (100) surface are presented in Table \ref{ss100}. Shown are the total adsorption energy, $E_{tot}$, in eV, the average adsorption energy $E_{av}$ per atom, and the distances $\delta d_{upp} $, $\delta d_{bot} $, between the adsorbed atoms in the upper and lower (bottom) atomic layers and the surface plane of the  Pd crystal in units of $a/2$. Eneregies obtained using PBEsol functional are shown in parantheses. As in the case of the (111) surface, in addition to filling available tetrahedral and octahedral sites, there is a segregation site above the surface layer, shown as a pore state in the Table \ref{ss100}.  This state has an enlarged adsorption energy E$_{ss, pore}$= -0.72 eV, compared to that for the "hollow" site at the (100) surface  (-0.47 eV), and is shifted by a small distance $\delta d=0.2$ \AA \ above the surface plane.  Calculations using PBEsol functional gave a smaller value E$_{ss, pore}$= -0.62 eV, but still 0.15 eV larger than for the "hollow" site at the clean surface. The enlargement of the adsorption energy can be interpreted in terms of "coordination number" for the segregated H atoms. It was noticed in \cite{Nazarov,Lischka} that H binding energy is regularly enlarged  for the sites with a smaller coordination number, almost inversely proportional to it. So one could anticipate the increase of adsorption energy up to -0.6 eV.  

The adsorption energies for the sites below the surface are much closer to the pattern for the surface vacancy, e.g., the adsorption energy for the octahedral bottom state is E$_{ss, bot}$= -0.33 eV, very close to that for the surface vacancy. For segregation in the $n$H-V$_{ss}$ clusters  with n=2,4,6  H atoms, one of which is the pore state and the second is at the vacancy bottom site, the adsorption energy is close to a sum of separate contributions;  the positions of the H atoms at the pore site and the bottom are unchanged. In all such clusters H at the pore site acquires enlarged adsorption energy. Absorption energy for such clusters with 2 and 6 H atoms, calculated using PBE-sol functional (shown in  parentheses) give a somewhat smaller, but still noticeably enlarged adsorption energy.

The segregation of H atoms among tetrahedral sites results in a smaller average adsorption energy, see Table \ref{ss100} so that they will not be competitive in the subsequent H segregation. The total adsorption energy for any  $n$H-V$_{ss}$ cluster with $n \ge 4$ is higher than the subsurface vacancy formation energy, so that these clusters should be stable and will have high concentration in the thermal equilibrium state.
\begin{figure*}[tb]
\centering
		\subfigure[]{{\includegraphics[width=4.6cm]{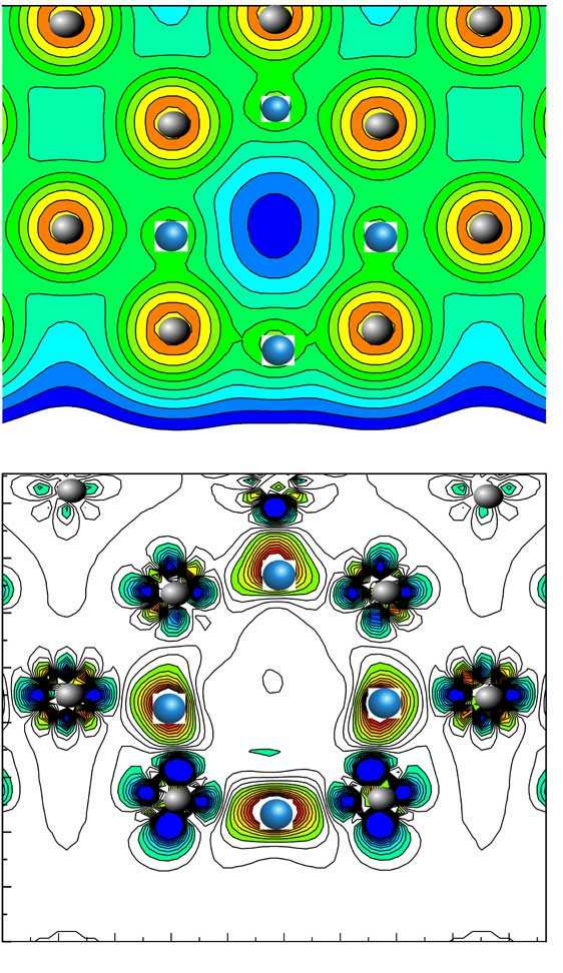} }}%
				\qquad
    \subfigure[]{{\includegraphics[width=6.3cm]{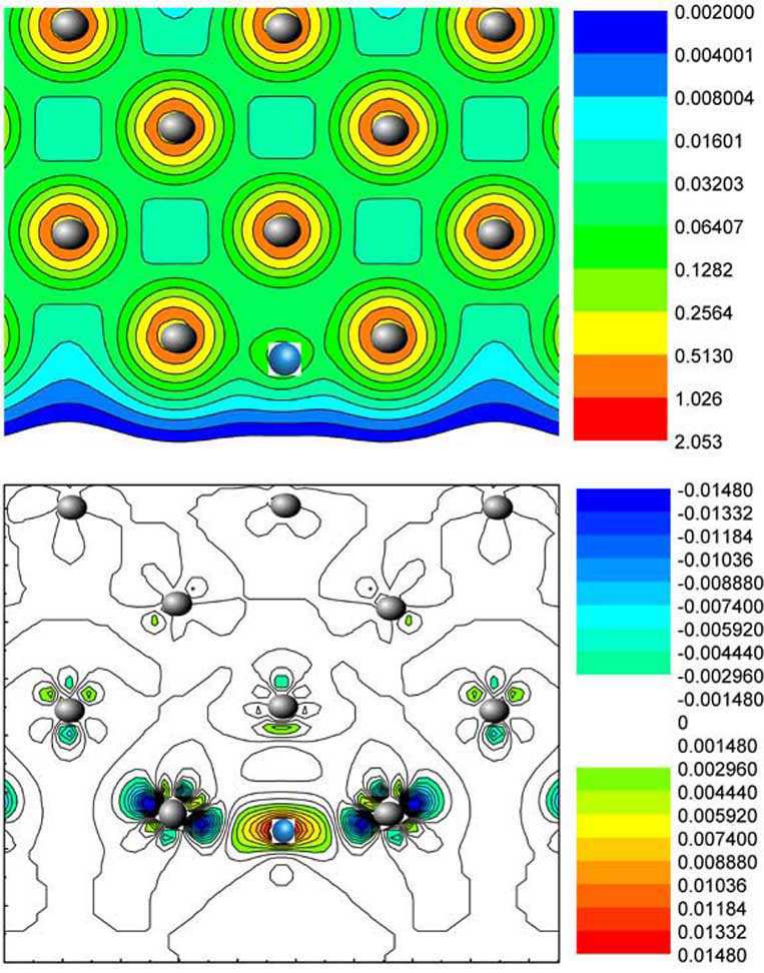} }}%
\caption[]{\label{4Hat100ssbv} (color online) (a) Contour plots of the charge distribution for 4 H atoms segregated in the 4H-V$_{ss}$ cluster at (100) Pd surface;  3 of 4 hydrogen  atoms (blue balls) are close to octahedral sites, while 1 H atom is shifted above the surface plane, forming a "pore" state with enlarged binding energy (upper panel); also shown is the charge redistribution for this cluster (lower panel). (b) Charge distribution in the plane of 1 H atom  segregated to Pd (100) surface with z axis normal to the surface; lower panel shows charge transfer from Pd to H atom.
Charge distribution notations are as in Fig. \ref{3H-Rhoat100surv}.}
\end{figure*}

The electron charge distribution for the 4H-V$_{ss}$ cluster at the Pd (100)  surface in the plane of segregated 4 H atoms  is shown in Fig. \ref{4Hat100ssbv} (a). Clearly seen is a displacement of the bottom site H atom below the 3-rd layer and the shift of subsurface atoms closer to the surface. A displacement in the subsurface plane  farther from the vacancy site by 0.3 \  \AA  ~ is also observed in the Fig. \ref{4Hat100ssbv} (a). For visual comparison we present the charge distribution for 1 H atom adsorbed to a hollow site at the Pd (100) surface, see Fig. \ref{4Hat100ssbv} (b). One can observe smaller lattice distortions and a closer placement of the H atom to the Pd surface. 

A charge density difference contour plots  $\delta \rho = \rho({\rm 31 Pd+4H})-\rho({\rm 31 Pd})-\rho({\rm 4H})$ for the same structures are presented in  Fig. \ref{4Hat100ssbv} (a),(b), lower panels. 
Charge accumulation and charge depletion contours are drawn from -0.1 to 0.1 e/\AA$^3$ with an interval 0.01 e/\AA$^3$. 
One can observe a net charge transfer from the adjacent $d$-orbitals of Pd to H atom, typical for H segregation in transition metals. However, (i) for all adsorption sites the charge transfer is at least 4 times smaller than for H segregated at vacancies and divacancies in  Ni \cite{Subashiev} or adsorbed at the Ni surface \cite{Kresse}; (ii) there is no preferential depletion of $d_z$2 orbitals as in Ni or in case of the  Pd (210) surface \cite{Lischka}. The main distinctive feature is a noticeably higher charge transfer for the pore state H atom, than for a hollow site of the Pd (100) surface. Apparently, this contributes to the enlargement of the adsorption energy for this state.  

The elementary processes of H segregation to $n$H-V$_s$ and to $n$H-V$_{ss}$ clusters at free Pd surfaces considered above can be observed only at low occupation of the surface sites by H atoms, $\theta \ll 1$. According to Eqs. (\ref{entha},\ref{mu}) it corresponds to $E_S-\mu_{H_2}/2 > k_BT$, i.e. a large and negative $\mu_{H_2}$ or a low ambient H$_2$ pressure \cite{Behm,Johansson}. When the amount of H on the surface is low, the vacancy concentration is exponentially small and does not change with H pressure. The concentration of the $n$H-V$_s$ clusters with segregated H is even much smaller because of much higher formation enthalpy. 

\subsection{Adsorption of H  on surface and subsurface vacancies at Pd surfaces covered with H monolayer}

Comparison of the data of previous sections for adsorption energies at the surface and at various cluster sites shows higher values of adsorption energy to the surface than the average segregation energy to the vacancies. Thus the adsorption to the surface sites prevails in all cases, except for the "above the pore" state of H at subsurface vacancies.

At higher H pressure (when $E_S-\mu_{H_2}(P,T)/2 < 0$) the hydrogen  atoms primarily fill the surface sites. For $\theta \rightarrow 1$ the formation energies for $n$H-V$_s$ and $n$H-V$_{ss}$ clusters should be calculated taking into account the high coverage of the Pd surface. The results of our calculations of the vacancy formation energies for the surface and subsurface layers of Pd(100) and adsorption energies for the slab surfaces at $\theta = 1$ (using PBEsol functional) are summarized in the Table \ref{Covered}. To eliminate possible effects of the electric dipole layer formation at the surface we tested all configurations for the 8 hollow sites on one surface of the slab occupied by H (starting with 8H$_s$-V$_s$ cluster, where subscript $s$ indicates H at the surface) and on both surfaces (i.e. 16H$_s$-V$_s$). The difference between the results in these two cases for the vacancy  formation energies and the adsorption energies were negligible. It corroborates with a small charge transfer to H atoms at the surface, seen in Fig.  \ref{4Hat100ssbv}.
We have also found that for $n$H-V$_{s}$ clusters the filling of free surface sites by H does not change considerably the positions of H and Pd atoms surrounding the vacancy compared to Fig. \ref{5Hat100surv}. Moreover, the adsorption energy per H$_s$ atom is $E_{ad}$=-0.5 eV for H atoms at the surface hollow sites. I.e. it remains close to the value for 1 H$_s$ on the clean Pd(100) surface ($E_{ad}$=-0.47 eV). This result as well as our calculation for the H adsorption energy at a 32-at slab without vacancies agrees with the data of Ref. \cite{Jung} but differs from some previous reports of the reduction of this energy  to $E_{ad}$=-0.3 eV \cite{Pallassana} for Pd(111) surface with the increase of $\theta$. The experimental data \cite{Johansson}  show a considerable decrease of the apparent heat of the adsorption with coverage which however was attributed by the authors to a variation of the sticking probability of H to the surface. 

The further filling of the bottom site by a H atom to result in 5H-V$_{s}$ cluster gives the adsorption energy $E_{bot}$=-0.31 eV for this site which is close to the former result for the 5H-V$_{s}$ cluster at a clean Pd(100) surface. Therefore the formation energy of 16H$_s$+1H-V$_s$ cluster on the H-covered surface is reduced by the $E_{bot}$ value as compared to the 16H$_s$-V$_{s}$ formation energy. Note, that this is almost half of the surface vacancy creation energy, so that the reduction is sizable.

Now we consider the effect of the surface coverage on the H adsorption to the $n$H-V$_{ss}$ clusters at Pd(100) surface.  First, our calculation showed that the vacancy formation energy in the P(100) subsurface layer $E_v=1.2$ eV is $0.05$ eV lower  for $\theta \approx 1$ then for for $\theta =0$ apparently because of the filling of the pore site. It is still 0.15 eV lower than for the bulk vacancy. Calculated total adsorption energy  atom for the 16H$_s$-V$_{ss}$ cluster at the surface is $E_{tot}$=-8.11 eV  exceeding that for the clean Pd surface. Note that one of H$_s$ atoms is at the "above the pore" site. So, the adsorption energy for this site is $E_{pore}$=-0.61 eV, close to -0.62 eV for the clean surface. At $\theta = 1$ the subsurface vacancy still has 5 octa-sites available for segregation. The segregation energies remain close to that for the  $n$H-V$_{ss}$ clusters at the free surface (with an average $E_{av}$=-0.3 eV). As before, the bottom site has the smallest adsorption energy, and the sites in the subsurface layer of 4H-V$_{ss}$ have  higher energy, $E_{av}$=-0.29 eV per H atom but the difference is small, see Table \ref{Covered}. The pore site is now shared by the vacancy and the surface hydrogen and therefore does not contribute to the energy gain in H segregation. The calculated total segregation energy $E_{tot}$=-1.31 eV exceeds the subsurface vacancy formation energy and should cause the increase of the $n$H-V$_{ss}$ cluster concentration at equilibrium. 

We have also studied the possibility of hydrogen molecule adsorption to the surface and subsurface vacancies, which can be facilitated at high ambient pressure. We have calculated energies $n$H-V$_{s}$ 
$n$H-V$_{ss}$ clusters in which a pair of H atoms were at the initial distance close to the distance in the H$_2$ molecule allowing relaxation to the ground state. We have found that in these clusters the  H$_2$ molecule dissociates and atoms find their place in their preferable sites. We have also found that in case of the high H coverage the excessive H atom in the vacancy is forced to the nearest bridge site at the surface. Similar effect was reported in modeling adsorption to the free Pd surface \cite{Jung}.

\section{Discussion} 
The distinct features of H segregation to surface and subsurface vacancies in Pd can be interpreted as a result of the effects of the short-range H interaction with the Pd electronic cloud and also the elastic surface relaxation effects. The configuration of H atoms at the (100) surface manifests a competition between segregation to the surface and segregation to a vacancy, which results in large shifts of the segregated H atoms from the vacancy site with the opposite shifts of Pd atoms. In all $n$H-V$_i$ clusters an average segregation energy per atom steadily decreases with the number of segregated H atoms.  
\begin{table}[t]
\begin{caption}{\label{Covered} Adsorption energies (in eV)  for H  to  $n$H-V$_{s}$ and $n$H-V$_{ss}$ clusters at Pd (100) surface and full coverage of the surface by H atoms in the hollow positions. Presented are vacancy creation energies,  total  adsorption energy for all H atoms, and the total energy of adsorption for the H atoms segregated to the vacancy in the cluster calculated using PBEsol functional. }
\end{caption}
\begin{tabular}{lllll}
\hline
 n & positions & $E_{v,s}$, $E_{v,ss}$  & $E_{tot}$ & $E_{t,vac}$   \\ \hline
16 & 16H$_s$-V$_{s}$     & 0.72 &  -8.02  & -   \\
17 & 16H$_s$, 1H-V$_{s}$ &  -  &  -8.31  &  -0.31 \\ 
16 & 16H$_s$-V$_{ss}$    & 1.2 &  -8.11  & - \\
17 & 16H$_s$, 1H-V$_{ss}$ & -  &  -8.39 & -0.28 \\ 
20 & 16H$_s$, 4H-V$_{ss}$ & -  &  -9.23 & -1.12 \\ 
21 & 16H$_s$, 5H-V$_{ss}$ & -  &  -9.42 & -1.31 \\ 
\hline
\end{tabular}
\end{table}

The large difference between the  formation energies for the vacancies at the surface, subsurface layers and the bulk suggests higher filling of the surface and subsurface layers with vacancies and also the variation of this concentration with the H adsorption. At low coverages and low H concentration in the Pd bulk the  surface and subsurface vacancies may act as the centers of H capture that alter the kinetics of H adsorption-desorption and H diffusion in the Pd bulk. 

At high $\theta \rightarrow 1$ due to the H segregation to available bottom site the concentration of 1H-V$_{s}$ clusters varies with the enthalpy of the formation of the defects, $H^f=E_{s}+E_{bot}-\mu_{H_2}(P,T)/2$.   The exponential increase of the vacancy concentration due H segregation and accumulation of 1H-V$_s$ clusters starts at the hydrogen pressure for which $E_{bot}-\mu_{H_2}(P,T)/2 \le 0 $. Even at the high H pressure, when $\mu_{H_2}(P,T) > 0 $ the formation enthalpy $H^f>0$ and the concentration increase is moderate. The reason is that the segregation sites surrounding the vacancy are shared with the surface hollow sites and no additional H sites are available for segregation. 

More important is the effect of ambient pressure on the concentration of the $n$H-V$_{ss}$ clusters. There are additional (to the above the pore) 5 sites available for segregation, and the calculated formation energy for the 5H-V$_{ss}$ cluster is negative, $E_f=E_v+E_{tot}$=-0.12 eV.  The cluster formation enthalpy $H^f=E_{ss}+5(E_{av}-\mu_{H_2}(P,T)/2)$ starts to drop down at $E_{av}-\mu_{H_2}(P,T)/2 \le 0 $, i.e. almost at the same pressure as in case of surface vacancy cluster. 

The increase in the cluster concentration at $T_0$=300 K and $P_0$ =1 bar due to the H segregation can be estimated as $ c_n/c_0=\exp\big\{-n[E_{av}-\mu_{H_2}(P_0,T_0)/2] /k_B T\big\} $, where $c_0$ is the surface/subsurface vacancy concentration in the pure Pd. For $E_{av}=-0.3$ eV it gives $c_1/c_0\approx 320$ and $c_5/c_0\approx 3.3 \times 10^{12}$. Thus, segregation should promote accumulation of $5$H-V$_{ss}$ clusters in the subsurface layer.  
Note that the estimation of the increase does not rely on the low accuracy evaluation of the vacancy formation energy. The steep decrease of 5H-V$_{ss}$ cluster formation enthalpy can result in instability of the system against the increase of the cluster concentration at the equilibrium P when $H^f(P,T) =0$. The increase in the  concentration will be limited by the interaction between the clusters and competition with possible formation of an ordered hydride-vacancy phase\cite{Zhang}. More accurate estimation of the cluster concentration should take into account the cluster formation volume and the configuration entropy. 

The same conclusions can be drawn for the $n$H-V$_{s}$ and $n$H-V$_{ss}$ clusters at Pd (111) surface at $\theta\rightarrow 1$: the H segregation to the tetrahedral subsurface sites of the surface vacancy eliminates the adsorption to the surface sites, while it does not affect segregation to the $n$H-V$_{ss}$ cluster. The detailed analysis for this case is more complicated because of more strong dependence of the surface adsorption energy on $\theta$. 

Similar mechanism is responsible for the increase of the vacancy cluster concentration in the bulk. As shown above the the threshold H pressure for the increase of the $n$H-V$_i$ cluster concentration is defined by $E_{av}^i-\mu_{H_2}(P,T)/2 =0 $ and it varies between the dominating for a given $i=s,ss,n$ clusters in a narrow range. More important for the  value of enthalpy $H^f$ above the threshold is the is the vacancy formation energy without H segregation. Therefore accumulation of $n$H-V$_{ss}$ clusters should prevail at these pressures. 

One can compare the H segregation in Pd with H adsorption to vacancies in Ni, the other fcc metal that exhibits SAV state formation \cite{Fukai_SAV}. The adsorption energies to the surface are very close for both metals \cite{Ferrin}, though the vacancy formation energy is higher in Ni (1.74 eV). The solubility of H is much higher in Pd, which in part is the result of the larger sizes of octa- and tetra-pores \cite{You}. The electronic charge redistribution accompanying H adsorption in Pd is much weaker than in $n$H-V and $n$H-V$_2$ clusters in nickel \cite{Subashiev,Kresse}. However in spite of seemingly much stronger Coulomb repulsion between the segregated H atoms, the average binding energy per atom for these clusters in Ni remains the same for n=1-6 while in Pd it steadily decreases with $n$. This should be attributed to much stronger effects of lattice relaxation in Pd, clearly evident in the $n$H-V$_{s}$ clusters at the (100) surface. Apparently, it is the result of larger (for Pd) lattice constant, smaller cohesive energy and larger bulk modulus. 

The other common feature in both metals is an almost doubled binding energy for H segregation to divacancies \cite{Nazarov,Subashiev,Tanguy}. It is observed experimentally and is attributed to the H segregation to octahedral sites in divacancies "shared" between them (having both vacancy sites as the nearest neighbors). The origin of the higher segregation energy is lower coordination number for the segregated H atom. Similar effect is found here for the H segregation to a pore state at the subsurface vacancy. Further comparison of the properties of surface and subsurface vacancies in Pd allows us to predict the above the pore adsorption sites for Ni which should have enlarged binding energy.

The common feature of all fcc metals is stronger binding of the adsorbed H to the surface than to vacancies \cite{Nazarov}. In addition, the adsorption energy of a single H atom to a vacancy never overcomes the surface vacancy formation energy. Therefore, because of sharing the segregation sites the surface layer will not be damaged by the excessive formation of $n$H-V$_{s}$ clusters. However there are 5 available free sites for segregation at the subsurface vacancy for (100) surface vacancy and 6 for the (111) subsurface vacancy (see Table \ref{subs111}) at the H-covered surface. The subsurface vacancy formation energy is smaller and the segregation energy is at least the same as in the bulk. Therefore the accumulation of $n$H-V$_{ss}$ clusters in the subsurface layer should be the common feature for all metals in which the SAV phase is observed. 
 
The barrier for H diffusion from surface to subsurface layers in Pd is known to be about 0.4-0.5 eV \cite{Ferrin,Mitsui}, almost 2 times higher than the diffusion activation energy in the bulk. The barriers for H atoms segregated in different sites at the surface is quite low, lower than 0.1 eV and the barrier for escaping from the subsurface vacancy into the bulk is less than from the surface. Therefore the $n$H-V$_s$ and $n$H-V$_{ss}$ clusters can alter kinetics of Pd charging with H. This conclusion is in line with experimental findings in electrochemical charging of Pd, accelerated when the surface vacancies are created at a high charging current and in H-Pd codeposition which produces a very high effective H pressure \cite{Benck}.

Vacancy formation in Pd bulk is known to be prompted by heating to high temperatures or by a rapid cooling. They can also be generated by mechanical and electrochemical treatments. In view of a high formation energy, the thermal activation of vacancies in the bulk with subsequent segregation of H atoms is an extremely slow process that cannot account for a SAV state formation and further ordering into a vacancy-rich Pd hydride phase. The formation energy of a vacancy in the surface or the subsurface layers is more effective because of the smaller formation energy.  In the ambient H$_2$ atmosphere the high concentration of surface hydrogen with much higher surface diffusion rates can promote formation of $n$H-V$_{s}$ and $n$H-V$_{ss}$clusters which can facilitate the SAV state formation. 

\section{Conclusions}
 We have studied the H segregation at the surface and subsurface vacancies at (111) and (100) surfaces of Pd using DFT calculations. The highest hydrogen adsorption energy to the cluster (-0.72 eV PBE, -0.62 PBEsol) is found for a pore state above the subsurface vacancy V$_{ss}$  in (100) plane. The high absorption energy for this state  manifests itself for all $n$H-V$_{ss}$ clusters with one of the H atoms bound to the pore site.  The binding energy remains high for up to 6 segregated H atoms. 
Enlarged binding energy is also found for the "pore" state  above the subsurface vacancy at the Pd (111) surface. For other configurations the binding energy is close to that of the bulk.  At high initial surface coverage by H the competition between the surface adsorption and adsorption to  $n$H-V$_{s}$ cluster limits the energy gain in H segregation, but does not affect segregation to $n$H-V$_{ss}$ clusters, resulting in a negative formation energy for $n$H-V$_{ss}$ clusters with $n \geq 4$.  Accumulation of the $n$H-V$_{ss}$ clusters should precede the formation of the superabundant vacancy phase. 

The authors are grateful to Paul Maye for valuable discussions and for helpful remarks to the text.

\setcounter{equation}{0}%
\setcounter{table}{0}%
\setcounter{figure}{0}%

\renewcommand{\theequation}{\arabic{equation}.S}
\renewcommand{\thetable}{\arabic{table}.S}
\renewcommand{\thefigure}{\arabic{figure}.S}

\section*{Supplemental Material}

We have calculated the whole set of bulk parameters of Pd: the fcc lattice parameter,  the cohesive energy, the bulk modulus, and the bulk vacancy formation energy, for LDA (PZ), GGA (PBE) and GGA (PBEsol) exchange-correlation functionals using the Quantum Espresso (QE) code and ultrasoft pseudopotentials available in QE library. 

The bulk vacancy formation energy was calculated using the formula 
\be
E^f(N)  = E_{t}(N-1)- \frac {N-1}{N} E_{t}(N)
\ee
where $E_{t}(N-1)$ is the ground state total energy of a $N$-site supercell with $N-1$ Pd atoms and a vacancy, and $E_{t}(N)$ is the energy for the pure Pd $N$-atom supercell. 

The results of the QE calculations and also the experimental results from Refs. \cite[]{Fukai,Ehrhart,Voter} and the calculation results of \cite[]{Nazarov,Csonka,Medasa,Sun,Vekilova} are summarized in Table \ref{TabNic0}. Tables with results of the other previous calculations can be found in e.g., \cite[]{Nazarov}. 
Results for a vacancy formation energy were obtained with the 31-at supercell and checked with the 107-at supercell. For fully relaxed lattice the formation energy is $\approx 10$ meV lower for a larger cell, which gives the estimate of concentration effects.

\begin{table*}[t] 
\begin{caption}{\label{TabNic0}
 Bulk properties of Pd: Comparison of the available {\em ab-initio} calculation and experimental results.}
\end{caption}
\begin{ruledtabular}
\begin{tabular}{c |c| c|c|c|c|c}    
           &  $a$,  a.u. (\AA)  &  $B$, GPa & $E_{coh}$, eV  &  $E_{vac}$ eV  & Ref. \\ \hline
  Exp                & 3.88   & 180 (195)   & 3.89 (3.94)  & 1.51, 1.7, 1.85 & \cite[]{Fukai,Ehrhart,Voter} \\
	QE  GGA (PBE)     & 3.93 & 198.8 & 3.79     &  1.2 &\\
	QE  GGA (PBE-sol)  & 3.87 & 192  & 4.41    &  1.34& \\
  QE LDA (PZ)        & 3.84 & 205   &  4.93  &  1.44 & \\
	VASP, GGA (PBE)   &3.94 (3.96)  & 169.4  &  3.69 (3.74)  & 1.19 (1.23) &\cite[]{Nazarov,Csonka,Sun}\\
  VASP  GGA (PBE-sol)  & 3.93 (3.87) & 186.0 (205) & 4.04 (4.47) & 1.45 & \cite[]{Csonka,Sun}\\
	VASP, LDA         & 3.86 & 206 (226) & 5.06 (5.96) & 1.48  &\cite[]{Nazarov,Vekilova,Csonka}\\
  VASP, meta-GGA, TPSS (revTPSS) & 3.89 (3.88) & 195 (206) & 3.98 (4.4) & - (1.71)  &\cite[]{Sun},(\cite[]{Medasa})\\
\end{tabular}
\end{ruledtabular}
\end{table*}

The results for the lattice constant are much less sensitive to the choice of  the exchange-correlation functional than the other parameters, especially the cohesive energy, and the elastic constant.  The experimental data for the vacancy formation energy have a large spread, and were obtained by different experimental techniques (specific heat, resistivity and positron annihilation measurements). The available results of numerous calculations using the same functionals also have some spread, caused by the differences in the pseudopotentials and in the sets of the calculation parameters. In the Table, presented are lower and upper bounds.  

One can see that the results obtained with TTPS functional match well with the experimental results making it the best choice. However the TTPS functional (and also its other versions like revTTPS) is not yet implemented in most of the solid state DFT codes except for VASP. The other problem is the convergence of the calculations with TTPS, which is known to be more difficult to achieve, especially for large supercells with low symmetry. The PBEsol functional gives the improved results compared to PBE or PZ functionals, and has no problems with the convergence. Hence it is the second best choice.

The surface energy per one atom at the surface for differently oriented slabs was evaluated for a periodic structure with a vacuum layer as
\be
E_{surf}=\frac{1}{2N_s }[E_{slab}(N_{sc})-N_{sc}E_{cry,p.a}] 
\label{SurfEn}
\ee
Here $N_s$ is the number of atoms on one surface  of a slab calculated using a supercell with $N_{sc}$ atoms in it,  $E_{cry,p.a}$ is the total energy per one atom in a bulk crystal calculated with exactly the same set of all parameters.  The vacancy formation energy for the slab is calculated by
\be
E_{vac,i}=[E_{slab,i}(N_{sc}-1)-E_{slab}(N_{sc}) + E_{cry,p.a}] 
\ee
Here index $i = s, ss$ for the surface and subsurface layer vacancies. 

The cut-off energies and the convergence threshold were chosen to achieve the bulk energy per atom scaling accuracy 2 meV  when going from 4-at to
27-at, 32-at, and further to 108-at supercell, (similar to Refs. \cite[]{Nazarov,NazarovFe}).  Therefore the accuracy of the energy calculation for the 36-atom supercell of the Pd(111) slab is about 70 meV in case of the total lattice relaxation (zero residual stress and forces per atom).  Using larger supercells results in a decrease of accuracy, together with a much larger calculation time. Smaller supercells will result in the increased interaction between the clusters.  The accuracy in calculating the differences in energies should be $\pm 140$ meV.  Expecting partial compensation of the regular variations in the differences of the total energy values one can estimate the calculation accuracy of about 0.1 eV.
\begin{figure}[b]%
 \includegraphics[width=7.2cm]{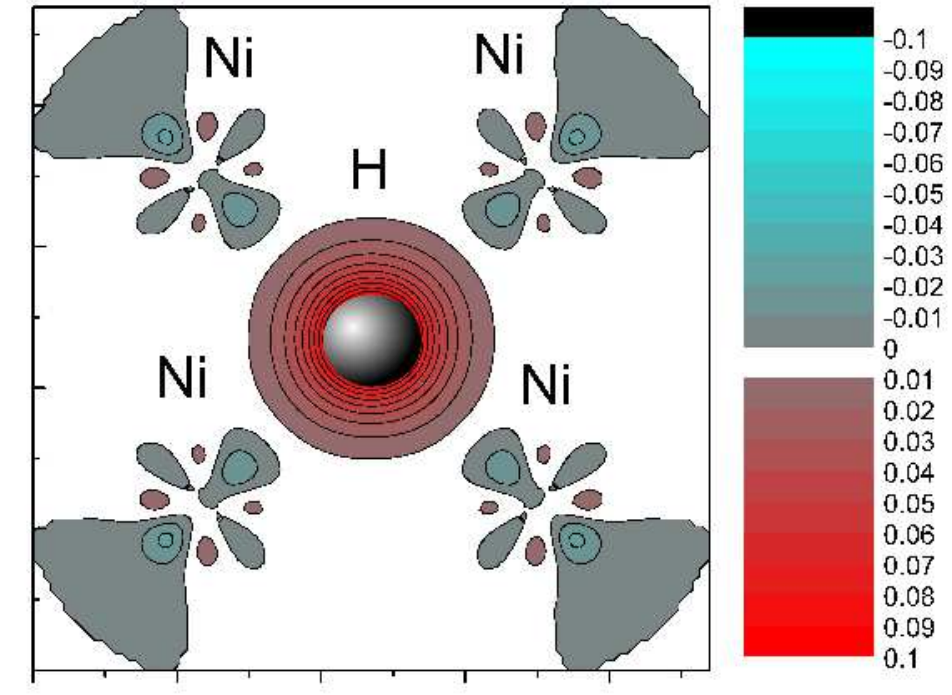}
    \caption{\label{Ni-toH-ch}Charge transfer for H interstitial in octa-site of Ni. Blue contours indicate the electron charge depletion, pink and red indicate charge accumulation. Charge density contours are drawn from -0.1, to 0.1 with the interval of 0.01 atomic unuts (6.748 electrons/\AA$^3$).}%
\end{figure}

For calculation of Zero Point Energy (ZPE) corrections through the vibrational frequencies of the intersitials and segregated to vacancy H atoms we used available PHonon module of Quantum Espresso  distribution. To  reduce the size of dynamic matrix, we have calculated the phonon energies using small atomic clusters, e.g.,  4-at (for H atom at the fcc hollow site at the  Pd (111) surface) and 5-at supercell (for H interstitial) in the octahedral site (1/2,1/2,1/2) and tetrahedral (1/4,1/4,1/4) site with adjusted lattice parameter. The low-frequency vibrations with frequencies below 300 cm$^{-1}$ were interpreted as crystal vibrations. Triple-degenerate highest frequency was interpreted as the frequency of H local vibrations and were used to calculate zero-point energies of hydrogen vibration. This approach is justified by the fact that H local vibrations are highly localized. The vibration energies are 0.136 eV for the surface,  0.15 eV  for octa and 0.18 eV for tetra-sites. The obtained values appeared to be close to these calculated with ``frozen Pd lattice'' in \cite[]{You}.  Large-size supercell with H$_2$ molecule was used to calculate ZPE for H$_2$, $E_{ZPE, H_2}$= 130 meV per atom.

In Fig. \ref{Ni-toH-ch}  the redistribution of the charge density $\delta \rho=\rho_{\rm Ni+H}-\rho_{\rm Ni}-\rho_{\rm H}$ around the interstitial H atom at octa-site of Ni lattice is depicted.  The net transfer of the electron charge  from Ni  to H seen in  in Fig. \ref{Ni-toH-ch} is  much larger than that for any site in Pd. In Ni it is accompanied by the depletion of adjacent Ni $d_z2$ orbitals and results in a weak ionic Ni-H interaction. 



\begin{thebibliography}{79}
\bibitem{PP} In Figs. \ref{SurVac111-3h3},\ref{5Hat100surv}, to give a sense of perspective, the atoms in the second and the third layers are depicted slightly enlarged.
\end{thebibliography}
\end{document}